\documentclass[useAMS,usenatbib]{mn2e}
\usepackage{graphics,rotating,multirow,bm,color,hyperref,amsmath}
\usepackage{subfig}
\hypersetup{
    colorlinks=true,
    linkcolor=black,
    citecolor=black,
}
\usepackage{dsfont}
\usepackage{amssymb}
\usepackage{graphicx}
\usepackage{verbatim}
\usepackage{natbib}
\usepackage{eucal}
\usepackage{calligra}

\usepackage{amsfonts}
\usepackage{color}
\usepackage[normalem]{ulem}
\usepackage[T1]{fontenc}
\DeclareMathAlphabet{\mathscr}{OT1}{pzc}%
                                 {m}{it}

\usepackage{times,epstopdf}
\usepackage{epsfig}
\usepackage[usenames,dvipsnames]{xcolor}
\usepackage{url}

\title[Haloes and subhaloes in $f(R)$ gravity]
{Exploring the liminality: properties of haloes and subhaloes in borderline $f(R)$ gravity}
\author[Shi et.~al.]
  {Difu Shi$^{1}$\thanks{difu.shi@durham.ac.uk}, Baojiu~Li$^{1}$, Jiaxin~Han$^{1}$, Liang Gao$^{2,1}$, Wojciech~A.~Hellwing$^{1}$\\
  $^1$Institute for Computational Cosmology, Department of Physics, Durham University, South Road, Durham DH1 3LE, UK\\
  $^2$National Astronomical Observatories, Chinese Academy of Sciences, 20A Datun Road, Chaoyang District, Beijing 100012, China}

\pagerange{\pageref{firstpage}--\pageref{lastpage}} \pubyear{2015}

\def\LaTeX{L\kern-.36em\raise.3ex\hbox{a}\kern-.15em
    T\kern-.1667em\lower.7ex\hbox{E}\kern-.125emX}

\newcommand{\tcr}[1]{#1}
\newcommand{\tcred}[1]{#1}

\newcommand{\be}{\begin{eqnarray}}
\newcommand{\ee}{\end{eqnarray}}
\newcommand{\bes}{\begin{equation*}}
\newcommand{\ees}{\end{equation*}}
\newcommand{\bea}{\begin{eqnarray}}
\newcommand{\eea}{\end{eqnarray}}
\newcommand{\beas}{\begin{eqnarray*}}
\newcommand{\eeas}{\end{eqnarray*}}

\include{aas_journals}

\begin{document}

\label{firstpage}

\maketitle

\begin{abstract}
We investigate the properties of dark matter haloes and subhaloes in an $f(R)$ gravity model with $|f_{R0}|=10^{-6}$, using a very high-resolution N-body simulation. The model is a borderline between being cosmologically interesting and {yet still consistent with current data.}
We find that the halo mass function in this model has a maximum 20\% enhancement compared with the $\Lambda$CDM predictions between $z=1$ and $z=0$. Because of the chameleon mechanism which screens the deviation from standard gravity in dense environments, haloes more massive than $10^{13}h^{-1}M_\odot$ in this $f(R)$ model have very similar properties to haloes of similar mass in $\Lambda$CDM, while less massive haloes, such as that of the Milky Way, can have steeper inner density profiles and higher velocity dispersions due to their weaker screening. The halo concentration is remarkably enhanced for low-mass haloes in this model due to a deepening of the total gravitational potential. Contrary to the naive expectation, the halo formation time $z_f$ is {\it later} for low-mass haloes in this model, a consequence of these haloes growing faster than their counterparts in $\Lambda$CDM at late times and the definition of $z_f$.
Subhaloes, especially those less massive than $10^{11}h^{-1}M_\odot$, are substantially more abundant in this $f(R)$ model for host haloes less massive than $10^{13}h^{-1}M_\odot$. 
We discuss the implications of these results for the Milky Way satellite abundance problem. 
Although the overall halo and subhalo properties in this borderline $f(R)$ model are close to their $\Lambda$CDM predictions, our results suggest that studies of the Local Group and astrophysical systems, aided by high-resolution simulations, 
{can be valuable for further tests of it.}
\end{abstract}


\section{Introduction}
\label{sect:intro}

The observed accelerated cosmic expansion is one of the most puzzling problems in modern physics \citep[e.g.,][]{weinberg}. In less than twenty years, it has motivated the proposal of a huge number of models. Apart from the current standard $\Lambda$CDM paradigm, in which the acceleration is driven by a cosmological constant $\Lambda$, such models are divided roughly in two classes. The first class introduces new physics in the particle sector and suggests that the acceleration is due to some new matter species, often known as dark energy. The second class proposes new physics in the gravity sector, so that the standard theory of gravity, Einstein's General Relativity (GR), is modified on cosmological scales to accommodate the accelerated expansion. This latter class of theories are commonly referred to as modified gravity \citep{clifton,Joyce2014} which is increasingly becoming an active research area.

For over a decade, the well-known $f(R)$ gravity \citep[][]{Carroll2003, Carroll2004} model has been a leading modified gravity candidate to explain the cosmic acceleration, although it actually has a much longer history in other contexts. It is a subclass of the more general theory called the chameleon theory \citep{Khoury2004}, in which an extra scalar degree of freedom is invoked, that can mediate a modification to the standard gravitational force of GR (known as the fifth force). This deviation from GR is not necessarily ruled out by local experiments, as the theory can employ the chameleon mechanism \citep{Khoury2004} to suppress the fifth force in dense environments such as the Solar System (see below for more details about this so-called chameleon screening). This means that the theory could pass local gravity tests. However, in less dense environments, such as those encountered on cosmological scales, the deviation from GR becomes sizeable, which means that cosmology can provide a unique means to probe new physics of this kind.

There are several important features of $f(R)$ gravity, some of which seem to have not been emphasised enough. Firstly, it is a well accepted perception that $f(R)$ gravity is flexible and, thanks to its 4th-order field equations, can, in principle, accommodate arbitrary background cosmologies \citep[see, for instance,][for a concrete example of $\Lambda$CDM background cosmology]{He2013}. In spite of the general impression that `$f(R)$ gravity can accelerate the cosmic expansion', it should be noticed that there is no necessary connection between the `acceleration' and `modified gravity' parts of $f(R)$ gravity: the well-studied model of \citet{hs2007}, as an example, can essentially be written as a cosmological constant {\it plus} a modification to the GR gravitational law. In this sense, $f(R)$ gravity is not a `better' model than $\Lambda$CDM, but rather one with a different nature of gravity, the study of which can shed light on the question why GR is successful and whether cosmological data can disapprove it.

Secondly, the chameleon screening in $f(R)$ gravity is a mechanism that `could work', but not necessarily `will work' -- whether it works depends on the system under consideration and the functional form of $f(R)$. Again, taking the \citet{hs2007} model for example: how efficiently the screening works is determined by a model parameter $|f_{R0}|$ (and another parameter $n$ which is often fixed; see below). Increasing this parameter makes it less likely for the screening to be effective. Values of $|f_{R0}|\lesssim10^{-6}$ are more difficult to distinguish from GR using cosmological observations, implying a limit of cosmological constraints. On the other hand, there are recent claims that $|f_{R0}|=10^{-6}$ could be in tension with astrophysical observations \citep[see, e.g.,][]{bvj2013}. Given that the chameleon mechanism works with different efficiency in different environments, it is critical to examine whether these stringent constraints become weaker when the environments of the astrophysical systems are more accurately modelled. The same could be said about terrestrial tests of the chameleon theory \citep[see, e.g.,][for references to some pioneering works in this direction]{bbdms2007a, bbdms2007b}.

For this reason, the \citet{hs2007} $f(R)$ gravity model with $|f_{R0}|=10^{-6}$ could be considered as being {\it borderline} between cosmological and {astrophysical constraints}: for higher values the model will probably have trouble with local and astrophysical tests, and for lower values the model is likely to be no longer interesting cosmologically. Here, we suggest a `bisection' approach to the study of $f(R)$ gravity: we first conduct a detailed investigation of the cosmological and astrophysical implications of the model with $|f_{R0}|=10^{-6}$, and then push the study and the resulting constraints to larger or smaller values based on the outcome. We hope to use this paper, in which we will concentrate on the cosmological aspects,  as an initial step in this direction, to motivate further, more in-depth, studies.

In this work, we employ one of the highest-resolution N-body simulations of $f(R)$ gravity currently available to study its effects on the properties of dark matter haloes and their subhaloes. These are the fundamental building blocks of the large-scale structure of our Universe and are closely connected with cosmological observations such as galaxy surveys. Previous studies have shown that the model considered here makes rather similar predictions to GR for many other cosmological observables, such as the matter and velocity power spectra \citep{LiPS2013, Hellwing2013, Gongbo-halofit, Taruya2014}, void properties \citep{Cai2014, Zivick2014}, redshift space distortions \citep{Jennings2012}, the integrated Sachs Wolfe effects \citep{CaifR2014} and X-ray scaling relations of clusters \citep{aps2013}, but the simulation resolution used have not been high enough to study haloes and subhaloes in great detail \citep[see, e.g.,][for a recent high-resolution zoom-in simulation which has a different focus from that of this paper]{ctl2015}.

This paper is structured as following: In \S\ref{sect:theory} we very briefly describe the $f(R)$ model studied here and summarise the technical specifications of our simulations. \S\ref{sect:halos} and \S\ref{sect:subhalos} present our detailed analyses of halo and subhalo properties respectively, and comparisons with the $\Lambda$CDM model. Finally, we summarise and conclude in \S\ref{sect:summary}.

Throughout this paper, we use the unit $c=1$, where $c$ is the speed of light.

\section{\lowercase{$f$}$(R)$ gravity and simulations}
\label{sect:theory}

In this section we briefly review the general theory of $f(R)$ gravity (\S\ref{subsect:MG_theory}), motivate the model which we focus on (\S\ref{subsect:fR_theory}) and describe the algorithm and technical specifications of our cosmological simulations (\S\ref{subsect:simulation}).

\subsection{$f(R)$ gravity and chameleon screening}
\label{subsect:MG_theory}

The $f(R)$ gravity model is designed as an alternative to dark energy to explain the accelerated expansion of the Universe. It generalises the Ricci scalar $R$ to a function of $R$ in the Einstein-Hilbert action,
\begin{eqnarray}
S &=& \int{\rm d}^4 x \sqrt{-g}~\frac{R+f(R)}{16\pi G},
\label{equ:E-H_action}
\end{eqnarray}
where $G$ is Newton's constant and $g$ is the determinant of the metric $g_{\mu\nu}$. 

Minimising the action Eq.~(\ref{equ:E-H_action}) with respect to the metric tensor $g_{\mu\nu}$ leads to the modified Einstein equation
\begin{eqnarray}
G_{\mu\nu}+f_{R}R_{\mu\nu}-g_{\mu\nu}\left[\frac{1}{2}f-\square f_{R}\right]-\nabla_\mu \nabla_\nu f(R)=8\pi G T^m_{\mu\nu},
\label{equ:E_equaiton}
\end{eqnarray}
where $G_{\mu\nu}$ is the Einstein tensor, $f_R\equiv{\rm d}f/{\rm d}R$, $\nabla_\mu$ is the covariant derivative, $\square\equiv\nabla^\alpha\nabla_\alpha$ and $T^m_{\mu\nu}$ the energy momentum tensor for matter fields. As $R$ contains second-order derivatives of $g_{\mu\nu}$, Eq.~(\ref{equ:E_equaiton}) has up to fourth-order derivatives. It is helpful to consider it as the standard Einstein equation for general relativity with an additional scalar field $f_R$. By taking the trace of Eq.~(\ref{equ:E_equaiton}), the equation of motion for $f_R$ can be obtained as
\begin{eqnarray}
\square f_R = \frac{1}{3}\left(R-f_R R+2f+8\pi G \rho_m\right),
\label{equ:fR}
\end{eqnarray}
where $\rho_m$ is matter density.

We consider a flat universe and focus on scales well below the horizon. On these scales, we can apply the quasi-static approximation by neglecting the time derivatives of $f_R$ in all field equations \citep[see, e.g.,][for tests which show that this approximation works well for the model studied here]{bhl2015}. Then Eq.~(\ref{equ:fR}) simplifies to
\begin{eqnarray}
\vec\nabla^2 f_R=-\frac{1}{3}a^2\left[R(f_R)-\bar R+8\pi G(\rho_m-\bar \rho_m)\right],
\label{equ:fR_no_T}
\end{eqnarray}
in which $\vec\nabla$ is the three-dimensional gradient operator and an overbar means we take the cosmological background value of a quantity. $a$ is the cosmic scale factor, normalised to $a=1$ today.
Similarly, the modified Poisson equation, which governs the Newtonian potential $\Phi$ in $f(R)$ gravity, can be simplified to
\begin{eqnarray}
\vec\nabla^2 \Phi=\frac{16\pi G}{3}a^2(\rho_m-\bar \rho_m)+\frac{1}{6}\left[R(f_R)-\bar R\right].
\label{equ:X}
\end{eqnarray}

%


There are two distinct regimes of solutions to the above equations:
\newline\indent$\bullet$ when $|f_R|\ll|\Phi|$, the GR solution $R=-8\pi G\rho_m$ holds to a good approximation and one has $\vec{\nabla}^2\Phi\approx4\pi G\delta\rho_m$ where we have defined $\delta\rho_m\equiv\rho_m-\bar{\rho}_m$, as the matter density perturbation. The effect of modified gravity is suppressed in this regime, which is a consequence of the scalar field being {\it screened} by the chameleon mechanism \citep{Khoury2004}.
\newline\indent$\bullet$ when $|f_R|\geq|\Phi|$, one has $|\delta R|\ll\delta\rho_m$ where $\delta R\equiv R-\bar{R}$, and so $\vec{\nabla}^2\Phi\approx16\pi G\delta\rho_m/3$. Compared with the standard Poisson equation in GR, we see a $1/3$ enhancement in the strength of gravity regardless of the functional form of $f(R)$. This is known as the {\it unscreened} regime, in which the chameleon mechanism does not work efficiently.

The chameleon mechanism is so named because it is most efficient in dense environments (or, more precisely speaking, regions of deep gravitational potential), where the scalar field $f_R$ acquires a heavy mass and the (Yukawa-type) modified gravitational force it mediates decays exponentially with distance so that it cannot be detected experimentally. The Solar System is one example of such an environment where $f(R)$ gravity {\it might} be in the screened regime and thus viable (i.e., not yet ruled out by local gravity experiment). However, to determine whether a specific $f(R)$ model is {\it indeed} viable is much more difficult, because this depends on the large-scale environments of the Solar System, such as the Milky Way Galaxy and its host dark matter halo. To assess this therefore requires high-resolution numerical simulations that can accurately describe these environments, and this is one goal of our paper. On the other hand, even if an $f(R)$ model passes local tests, there is still a possibility that it deviates significantly from GR on cosmic scales, where the chameleon mechanism is not as efficient. To study such deviations also requires accurate numerical simulations.

\subsection{The $f(R)$ model of this work}
\label{subsect:fR_theory}

In this work we study the model proposed by \citet[][hereafter HS]{hs2007}, which is specified by the following functional form of $f(R)$:
\begin{eqnarray}\label{eq:hs}
f(R) = -M^2\frac{c_1\left(-R/M^2\right)^n}{c_2\left(-R/M^2\right)^n+1},
\end{eqnarray}
in which $c_1, c_2$ are dimensionless model parameters, and $M^2\equiv8\pi G\bar{\rho}_{m0}/3=H_0^2\Omega_m$ is another parameter of mass dimension 2; here $H$ is the Hubble rate and $\Omega_m$ is the present-day matter energy density in units of the critical density ($\rho_c\equiv3H_0^2/8\pi G$). We always use a subscript $_0$ to denote the current value of a quantity unless otherwise stated.

When the background value of the Ricci scalar satisfies $|\bar{R}|\gg M^2$, we can simplify the trace of the modified Einstein equation of this model as
\begin{eqnarray}
-\bar{R} \approx 8\pi G\bar{\rho}_m-2\bar{f} \approx 3M^2\left(a^{-3}+\frac{2c_1}{3c_2}\right).
\end{eqnarray}
This is approximately what we have for the background cosmology in the standard $\Lambda$CDM model, with the following mapping
\begin{eqnarray}
\frac{c_1}{c_2} = 6\frac{\Omega_\Lambda}{\Omega_m},
\end{eqnarray}
where $\Omega_\Lambda\equiv1-\Omega_m$.

By taking $\Omega_\Lambda\approx0.7$ and $\Omega_m\approx0.3$, we have $|\bar{R}|\approx40M^2\gg M^2$ today (remember that $|\bar{R}|$ is even larger at earlier times), and so the above approximation works well.
Moreover, this can be used to further simplify the expression for $f_R$:
\begin{eqnarray}\label{eq:temp}
f_R \approx -n\frac{c_1}{c_2^2}\left(\frac{M^2}{-R}\right)^{n+1} < 0.
\end{eqnarray}
This can be easily inverted to obtain $R(f_R)$, which appears in the scalar field and modified Poisson equations as shown above.  As a result, two combinations of the three HS model parameters, namely $n$ and $c_1/c_2^2$, completely specify the model. In the literature, however, this model is often specified by $f_{R0}$ instead of $c_1/c_2^2$, because $f_{R0}$ has a more physical meaning (the value of the scalar field today), and the two are related by
\begin{eqnarray}
\frac{c_1}{c_2^2} = -\frac{1}{n}\left[3\left(1+4\frac{\Omega_\Lambda}{\Omega_m}\right)\right]^{n+1}f_{R0}.
\end{eqnarray}

We will focus on a particular HS $f(R)$ model with $n=1$ and $|f_{R0}|=10^{-6}$, which is sometimes also referred to in the literature as F6. Such a choice of $f_{R0}$ is made deliberately as a {\it borderline}: models with  $|f_{R0}|\geq10^{-5}$ are likely to already be in tension with cosmological observations \citep[see, e.g.,][for a review of current constraints on $f(R)$ gravity]{lombriser2014}, while those with $|f_{R0}|<10^{-6}$ are generally hard to distinguish from $\Lambda$CDM.


\subsection{Cosmological simulations of $f(R)$ gravity}
\label{subsect:simulation}


The simulation used in this work was executed using the {\sc Ecosmog} code \citep{ecosmog}. {\sc Ecosmog} is a modification to the publicly available N-body and hydro code {\sc Ramses} \citep{ramses}. New routines were added to solve the scalar field and modified Einstein equations in $f(R)$ gravity. This is a massively parallelised adaptive mesh refinement (AMR) code, which starts off from a uniform grid (the so-called domain grid) covering the cubic simulation box with $N^{1/3}_{\rm dc}$ cells on each side. When the effective particle number in a grid cell exceeds a pre-defined criterion ($N_{\rm ref}$), the cell is split into 8 daughter cells so that the code hierarchically achieves higher resolutions in dense environments. Such high resolutions are needed both to accurately trace the motion of particles and to ensure the accuracy of the fifth force solutions. The force resolution, $\epsilon_{\rm f}$, is twice the size of the cell which a particle is in, and we only quote the force resolution on the highest refinement level.

\begin{table}
\caption{The parameters and technical specifications of the N-body simulations of this work. 
$\epsilon_{\rm s}$ is the threshold value of the residual \citep[see, e.g.,][for a more detailed discussion]{LiPS2013} for the convergence of the scalar field solver. Note that $N_{\rm ref}$ is an array because we take different values at different refinement levels, and that $\sigma_8$ is for the $\Lambda$CDM model and only used to generate the initial conditions -- its value for $f(R)$ gravity is different but is irrelevant here.}
\begin{tabular}{@{}lll}
\hline\hline
parameter & physical meaning & value \\
\hline
$\Omega_m$  & present fractional matter density & $0.281$ \\
$\Omega_{\Lambda}$ & $1-\Omega_m$ & $0.719$ \\
\tcred{$\Omega_b$}  & \tcred{present fractional baryon density} & \tcred{$0.046$} \\
$h$ & $H_0/(100$~km~s$^{-1}$Mpc$^{-1})$ & $0.697$ \\
$n_s$ & primordial power spectral index & $0.971$ \\
$\sigma_{8}$ & r.m.s. linear density fluctuation & $0.820$ \\
\hline
$n$ & HS $f(R)$ parameter & $1.0$ \\
$f_{R0}$ & HS $f(R)$ parameter & $-1.0\times10^{-6}$ \\
\hline
$L_{\rm box}$ & simulation box size & 64~$h^{-1}$Mpc\\
$N_{\rm p}$ & simulation particle number & $512^3$\\
$m_{\rm p}$ & simulation particle mass & $1.52\times 10^{8}h^{-1}M_{\odot}$\\
$N_{\rm dc}$ & domain grid cell number & $512^3$\\
$N_{\rm ref}$ & refinement criterion & 3, 3, 3, 3, 4, 4, 4, 4...\\
$\epsilon_{\rm s}$ & scalar solver convergence criterion & $10^{-8}$ \\
$\epsilon_{\rm f}$ & force resolution & 1.95~$h^{-1}$kpc\\
\hline
$N_{\rm snap}$ & number of output snapshots & $122$ \\
$z_{\rm ini}$ & redshift when simulation starts & $49.0$ \\
$z_{\rm final}$ & redshift when simulation finishes & $0.0$ \\
\hline
\end{tabular}
\label{table:simulations}
\end{table}

\begin{figure*}
\includegraphics[width=18cm]{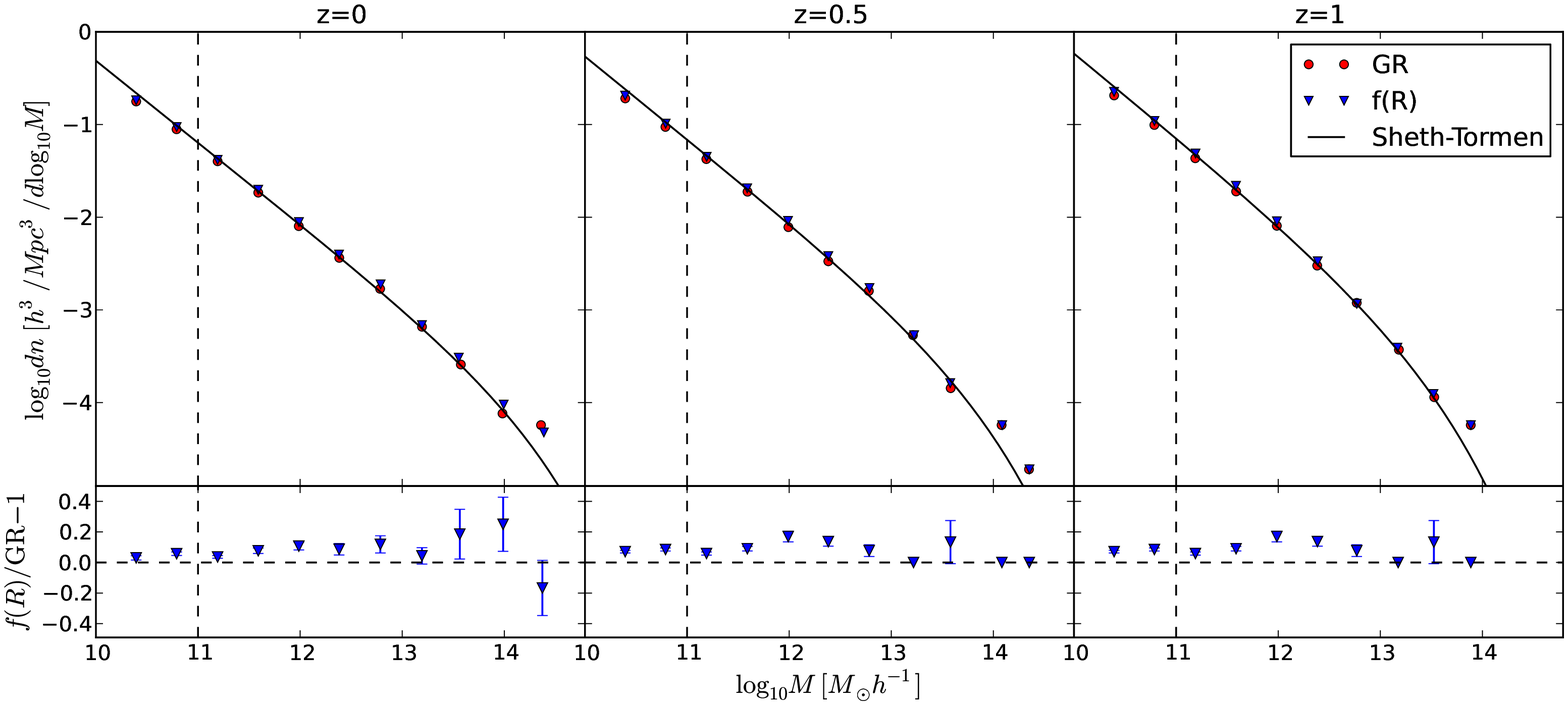}
\caption{Comparison of differential halo mass functions in GR (red circles) and F6 (blue triangles) with the \citet{Sheth1999} prediction for GR, at three redshifts -- $z=0.0$ (left panel), $0.5$ (middle panel) and $1.0$ (right panel). The relative difference between the two models is plotted in the bottom panels. Haloes are identified using a FoF algorithm with linking length 0.2. As we only have one realisation, the error bars are estimated from subsampling by dividing the simulation box into eight subboxes of equal size; the difference between F6 and GR mass functions, $\Delta n_i$, was computed for each subbox $i$, and its mean value ($\langle\Delta n\rangle$) and standard deviations ($\sigma_{\Delta n}$) were obtained using the values from the 8 subboxes; the relative difference was then calculated as $\langle\Delta n\rangle/\langle n_{\rm GR}\rangle$, with the error bars obtained in the standard way of error propagation.  \tcred{The vertical dashed line indicates a cut of our FoF halo catalogue at $\sim700$ particles, or $M_{\rm FoF}\sim10^{11} h^{-1}M_\odot$, for illustrative purpose, above which there is good agreement between the GR mass functions and the Sheth-Tormen fitting formulae (better than 10\%).}} 
\label{fig:hmf_FOF}
\end{figure*}

The parameters and technical specifications of our simulations are listed in Table~\ref{table:simulations}. The cosmological parameters are adopted from the best-fit $\Lambda$CDM cosmology of WMAP9 \citep{WMAP9}. The simulation was evolved from an initial redshift $z_{\rm ini}=49$ to today, and the initial conditions were generated using the {\sc Mpgrafic} package \citep{mpgrafic}. For comparison, we ran a $f(R)$ and a $\Lambda$CDM simulation using exactly the same initial conditions and the same technical specifications (we have used the $\Lambda$CDM initial conditions for the $f(R)$ gravity simulation because these two models are practically indistinguishable at epochs as early as $z_{\rm ini}=49$). \tcred{The small size of our simulation box implies that the properties of high-mass objects, such as their number densities, could be subject to run-by-run variations. However, the fact that our $f(R)$ and GR simulations start from the same initial conditions helps to suppress the run-by-run variation when we look at the relative difference between the predictions of the two models.}

With $512^3$ particles in a box of size $L_{\rm box}=64$$h^{-1}$Mpc, this is currently the highest resolution {\it cosmological} simulation of $f(R)$ gravity which runs to $z=0$. Another high-resolution $f(R)$ simulation has been conducted \citep{ctl2015}, in which the zoom-in technique was used to study the effects of $f(R)$ gravity on a Virgo-cluster-scale dark matter halo. Both simulations are purely dark matter. Recently, a  hydrodynamical simulation was carried out by \citet{aps2014}, which had a higher particle resolution and focused mainly on a different model parameter and early times, at which the model studied here is almost indistinguishable from GR.

\section{Properties of dark matter halos}
\label{sect:halos}

In this section we will concentrate on the properties of dark matter haloes measured from our simulations. Dark mater haloes are the most basic blocks of the large-scale structure and host the formation and evolution of galaxies. Therefore, the study of their properties is of great importance to the understanding of the fundamental nature of gravity. A number of halo properties have been studied in detail in the context of $f(R)$ gravity, such as the angular momentum, spin, velocity dispersion \citep{lzlk2013, hhlg2015}, velocity profile \citep{glmw2014} and screening \citep{zlk2011b, lzk2012, hlhg2014}. The improved resolution of our simulations enables us to study a wider range of the physical properties of haloes.

\subsection{Halo mass functions}

The differential mass function, ${\rm d}n(M, z)/{\rm d}\log M$, defined as the number of dark matter haloes 
per unit logarithmic mass found per unit volume, is an important theoretical and observational statistic of the dark matter density field. Indeed, the abundance of dark matter haloes is sensitive to the underlying cosmological model. Both N-body simulations and (semi-)analytical formulae have been used to predict the halo mass function \citep[see, e.g.,][for some examples of analytical mass function fitting formulae]{Sheth1999, jenkins, reed}.

In order to compare with the above-mentioned fitting formulae, we use the friends-of-friends (FoF) group-finding algorithm to identify dark matter haloes, using a linking length of $0.2$ times the mean inter-particle separation \citep{Davis1985}.

In Fig.~\ref{fig:hmf_FOF}, we plot the differential halo mass function measured from our simulations, along with the theoretical prediction for GR from \citet{Sheth1999} (upper panels), and the relative difference between $f(R)$ gravity and GR (lower panels) at $z=0$ (left), $0.5$ (middle) and $1.0$ (right). For the mass range we consider, the \citet{Sheth1999, jenkins} \& \citet{reed} fitting formulae all agree \tcr{reasonably} well and so we only plot one of them. We can see from the upper panels that the fitting formula describes very well the FOF halo mass function for GR at the redshifts studied, down to a halo mass of about $2-3\times10^{10}h^{-1}M_\odot$ (which corresponds to $\sim$200 simulation particles). The mismatch at masses above $\sim10^{14}h^{-1}M_\odot$ is due to the lack of volume for our small simulation box.


Fig.~\ref{fig:hmf_FOF} (lower panels) indicates that \tcr{the differential halo mass function for F6 model studied here is up to $\sim$20\% larger than the result for a} $\Lambda$CDM model with the same cosmological parameters. The difference is purely a result of the modified gravitational force in the F6 model. However, due to the strong chameleon screening in this model, the enhancement is very mild and hard to detect observationally. This is why we call F6 a borderline model -- it probably represents the limit achievable by many cosmological observations for the near future, even though it might still potentially be ruled out by {employing certain observables} \citep[e.g.,][]{fabian, zlk2011,Bel2014,lsm2015}, or using astrophysical observations \citep[e.g.][]{bvj2013}.

Inspecting the lower panels of Fig.~\ref{fig:hmf_FOF} more closely, we observe the trend that for very massive haloes, the mass functions for $f(R)$ gravity and GR agree, which is because the chameleon mechanism works efficiently for such haloes to suppress the effects of modified gravity. Disagreement between the two models appears below some critical mass, which increases with time, because at late times the chameleon mechanism is less efficient at suppressing modified gravity. Finally, at very low halo masses, we see the trend that GR starts to produce more haloes than $f(R)$ gravity, which is a result of a larger fraction of small haloes having been absorbed into big haloes in $f(R)$ gravity \citep{le2012}.

\begin{figure}
\includegraphics[width=8.8cm]{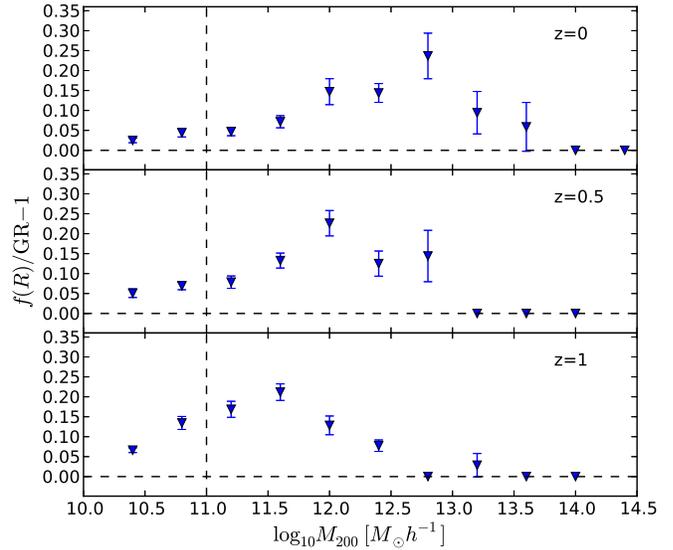}
\caption{The ratios of the differential halo mass functions between $f(R)$ gravity and GR, for the same three redshifts as in Fig.~\ref{fig:hmf_FOF}. Here the halo mass is $M_{200}$, defined as the mass within the radius at which the average density 200 times the critical density. The error bars are calculated in the same way as in Fig.~\ref{fig:hmf_FOF}. \tcred{The vertical dashed line indicates roughly the smallest halo mass ($700$ particles, or $M_{200}\sim10^{11} h^{-1}M_\odot$) we have used in the analyses of this paper.}}
\label{fig:hmf_M200}
\end{figure}

It is well known that certain properties of dark matter haloes, such as the mass function, depend on the halo definition used \citep[e.g.,][]{white2001,sawala}. In the above, to make comparison with the Sheth-Tormen formulae, we have used FOF haloes. When studying halo properties, what is more often used in the literature is $M_{200}$, the mass inside the radius $r_{200}$ within which the average density is 200 times the {\it critical} density, $\rho_c$. To check whether the choice of the halo definition affects our result, we plot in Fig.~\ref{fig:hmf_M200} the difference between the $f(R)$ and GR mass functions when using $M_{200}$, again at $z=0.0$ (upper panel), $0.5$ (middle) and $1.0$ (lower panel). We find the same qualitative features as in the lower panels of Fig.~\ref{fig:hmf_FOF}, but also some quantitative differences in the curves. In particular, the curves are smoother and better-behaved when using $M_{200}$, which may be because the FOF haloes are too irregular in their shapes and gravity is enhanced with different efficiency in different parts of the haloes, which can contaminate the screening effect expected for ideal spherical haloes \citep[see, e.g.,][for more discussion about the expected behaviour of the $f(R)$ halo mass function]{le2012,llam2012,llkz2013,lkl2014}.

\begin{figure*}
\includegraphics[width=17.8cm]{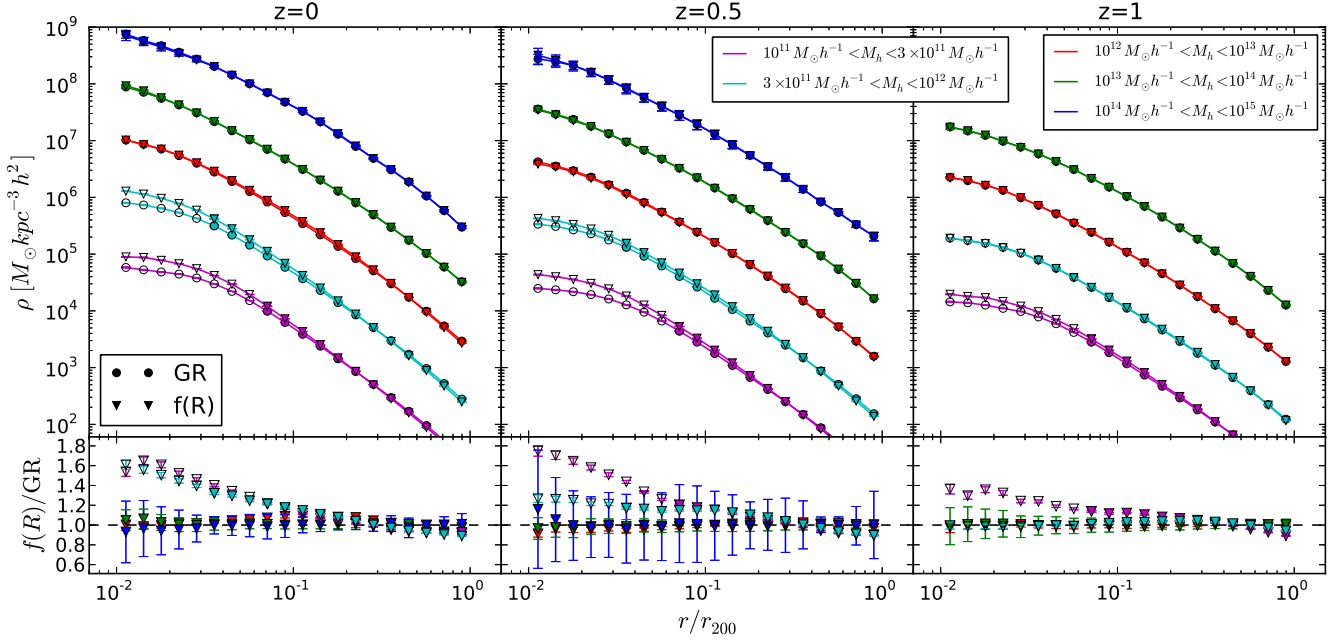}
\caption{Stacked dark matter halo density profiles for \tcr{five} mass bins (indicated by the legend) at three epochs: $z=0.0$ (left panel), $0.5$ (middle panel) and $1.0$ (right panel). {\it Upper panels}: density profiles -- results from the GR simulation are shown as circles while the $f(R)$ results are shown as triangles; the solid curves connect the symbols and the halo mass increases from the bottom curves to the top curves. {\it Lower panels}: the corresponding ratios between $f(R)$ gravity and GR which deviate from unity more for smaller mass bins. Open symbols are used when the distance from halo centre is smaller than the force resolution (which happens only in the lowest mass bin). The error bars are 1-$\sigma$ standard deviations for all haloes in each radius bin. To assist visualisation, in the top panels, we have rescaled the stacked density profiles of haloes within the mass bins \tcr{$10^{14}\sim 10^{15}h^{-1}M_\odot$, $10^{12}\sim 10^{13} h^{-1}M_\odot$, $3\times10^{11}\sim 10^{12} h^{-1}M_\odot$ and $10^{11}\sim 3\times10^{11} h^{-1}M_\odot$ by factors of $10$, $0.1$, $10^{-2}$ and $10^{-3}$ respectively. The numbers of haloes in each bin, starting from the most massive one, are respectively (numbers for the F6 simulation are in parentheses): 7 (7), 72 (78), 200 (232), 509 (586), 1558 (1714), 3758 (3936) at $z=0$, 3 (3), 62 (62), 194 (215), 539 (624), 1665 (1968), 4224 (4554) at $z=0.5$, and 0 (0), 48 (49), 152 (156), 508 (564), 1730 (2047), 4404 (5177) at $z=1.0$}.}
\label{fig:haloprof}
\end{figure*}

We will use $M_{200}$ in the rest of this paper, because of its wide use in the literature. Furthermore, to ensure good resolution of halo structure, we will \tcred{conservatively} restrict our analysis to haloes with more than 700 particles ($M_{200}\gtrsim10^{11} h^{-1}M_\odot$). \tcred{Even cut at $\sim$400 particles, we have found that the FoF mass functions at $z=0,0.5$ and $1$ show agreements with fitting formulae better than 10\%}.

\begin{figure*}
\includegraphics[width=18.1cm]{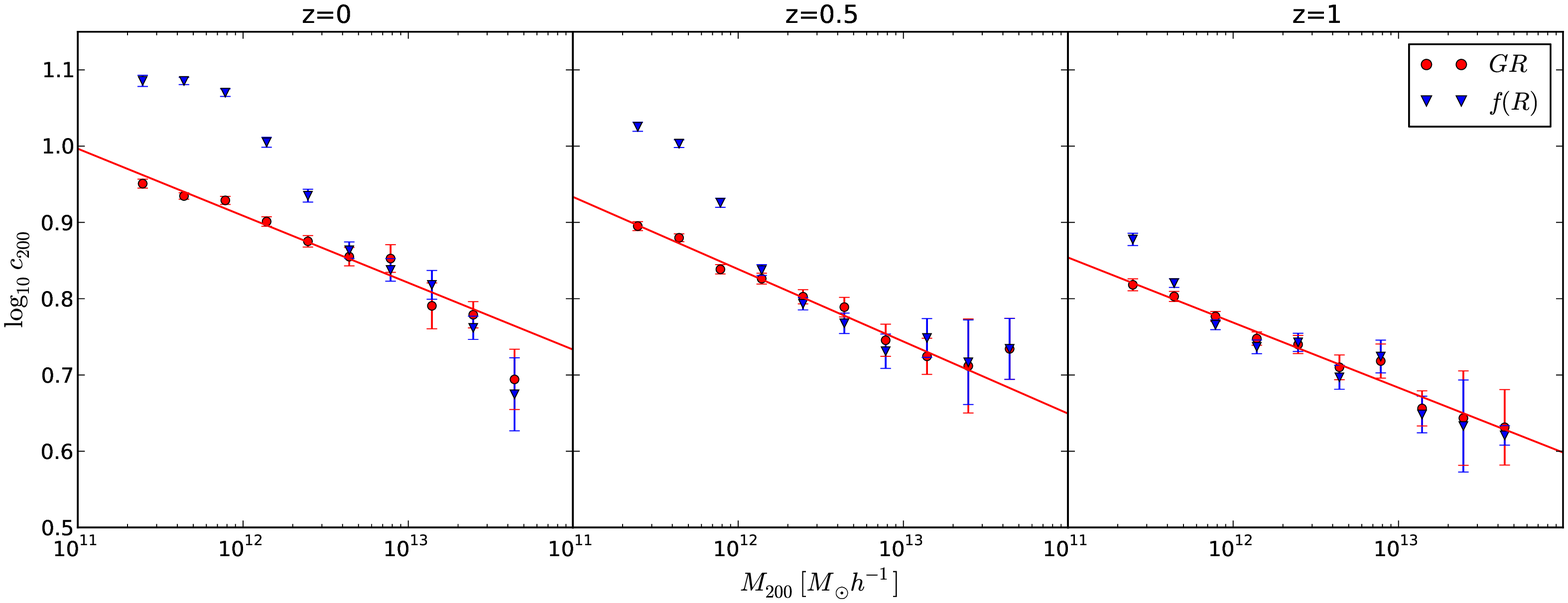}
\caption{Dark matter halo concentrations $c_{200}$ as a functions of $M_{200}$ at three redshifts, $z=0.0$ (left panel), $0.5$ (middle panel) and $1.0$ (right panel). Results for GR (F6) are shown using red circles (blue triangles), and the curves are power-law fits of the $c$-$M_{200}$ relations: in the case of GR (red solid line), the relation can be fitted using a single power law for the whole mass range, while for F6 this is no longer true \tcr{and so we do not show any fitting}. The error bars are obtained as the 1-$\sigma$ standard deviation of all haloes in each given mass bin.}
\label{fig:halo_c}
\end{figure*}

\subsection{Mass distribution inside haloes}

The inner structure of dark matter haloes provides invaluable information about their formation history, which can also be affected by the nature of gravity. In this subsection, we look at the dark matter density profiles and concentration-mass relations for haloes in the two models.

In Fig.~\ref{fig:haloprof}, we show the stacked halo dark matter density profiles at three redshifts $z=0.0$ (left), $0.5$ (middle) and $1.0$ (right). The distances from halo centres, as plotted on the horizontal axis, are rescaled by $r_{200}$, and all haloes with $M\geq10^{11}h^{-1}M_\odot$ in our simulations are divided into \tcr{5} mass ranges as indicated in the legend (the highest mass bin does not show up in the $z=1.0$ panel, since at that time the very massive haloes have not formed in great numbers). \tcr{Note that the widths of mass bins are different, and we do not make finer subdivisions of the three most massive bins since the model differences are small there, an observation we discuss now.}

From Fig.~\ref{fig:haloprof}, we see that the density profiles of haloes more massive than $10^{13}h^{-1}M_\odot$ show almost no difference between the two models at all three redshifts, because these haloes are very efficiently screened by the chameleon mechanism. The haloes in the mass bin $10^{11}\sim10^{12}h^{-1}M_\odot$ have up to 60\% higher density towards their centres in $f(R)$ gravity than in GR, because the screening efficiency is weaker. Thus, Milky-Way-sized haloes have steeper inner profiles in $f(R)$ gravity. Note, however, that as the force resolution of our simulations is $\sim 2$ $h^{-1}$kpc, 
we show the results within 5 times of it, i.e., $\sim 10$ $h^{-1}$kpc, using open rather than filled symbols. 
\tcr{We have explicitly checked that $10h^{-1}$ kpc is roughly equal to the Power convergence radius \citep{Power2003,Schaller2014} in our smallest halo mass bins $(10^{11}\sim10^{12}h^{-1}M_\odot)$, and is larger than the convergence radius for other halo mass bins shown in Fig.~\ref{fig:haloprof}. Though the Power radius is found by testing convergence on simulations with tree (and not AMR) codes, the physics of collisional relaxation used in its derivation is the same in our simulations, and so we use it as a reference.}
We conclude therefore that the region within $0.1\times r_{200}$ of Milky-Way-sized (and smaller) haloes in $f(R)$ gravity has a significantly steeper density profile than in GR, \tcr{and will further confirm this by studying the halo concentration-mass relations below.}

By comparing the three panels in Fig.~\ref{fig:haloprof}, it is also evident that the differences in the inner density profiles of the two models grow in time. This is as expected because the effect of modified gravity is cumulative, and also because at late times the chameleon screening is weaker in general, which leads to stronger modifications to GR. \tcr{In particular, haloes with masses below $\sim3\times10^{11}h^{-1}M_\odot$ already show signifiant discrepancy between F6 and GR at $z=1$, and for haloes from the mass bin $3\times10^{11}\sim10^{12}h^{-1}M_\odot$ the discrepancy starts at later times because of more efficient chameleon screening, although by $z=0$ the model differences have become roughly the same for these two bins.}

Next, we fit the dark matter density profiles in the two models using the Navarro-Frenk-White \citep[][NFW]{NFW96,NFW97} formula, which is given by
\begin{eqnarray}
{\rho(r)} = \frac{\rho_{\rm s}}{(r/r_{\rm s})(1+r/r_{\rm s})^2},
\label{equ:NFW}
\end{eqnarray}
in which $\rho_{\rm s}$ and $r_{\rm s}$ are the scale density and scale radius of the halo. The $\rho_{\rm s}$ and $r_{\rm s}$ parameters are connected to the halo mass, $M_{200}$ (or equivalently, the virial radius $r_{200}$), and concentration, $c$ (note that we have neglected the subscript in $c_{200}$ for brevity), through
\begin{eqnarray}
\rho_{\rm s} = \frac{200 \rho_{\rm c}}{3} \frac{c^3}{[\ln(1+c)-c/(1+c)]},
\label{equ:mass}
\end{eqnarray}
\begin{eqnarray}
c = r_{200}/r_{\rm s}.
\label{equ:c_def}
\end{eqnarray}
In practice, we obtain the $M_{200}$ and $r_{200}$ of each halo according to the spherical over-density definition, and estimate $c$ using Eq.~\eqref{equ:c_def} from the best-fitting $r_s$. \tcred{ \citet{llkz2013} found that haloes in $f(R)$ gravity can be well described by the NFW formula Eq.~(\ref{equ:NFW}). In this work, we have further confirmed this by explicitly checking the $\chi^2$ goodness-of-fit, in which we found that Eq.~(\ref{equ:NFW}) works almost equally well in GR and $f(R)$ gravity (with marginally smaller $\chi^2$ for haloes between $\sim10^{12}-10^{13}h^{-1}M_{\odot}$ in $f(R)$ gravity), though the  concentration parameters can be different, as we shall show below.}

Fig.~\ref{fig:halo_c} shows the halo concentration-mass relation, $c(M_{200})$, also at three redshifts $z=0.0$ (left), $0.5$ (middle) and $1.0$ (right), from which one can see clearly that the most massive haloes have nearly the same concentration in the two models, because the effects of modified gravity are efficiently screened in these objects. It is well known \tcr{from early studies} that the halo concentration in $\Lambda$CDM simulations is given by a power-law function of mass \citep[e.g.,][]{bullock,Zhao2003,neto,duffy,maccio,giocoli,dooley}, and our $\Lambda$CDM simulation shows the same result as illustrated by the red curves in Fig.~\ref{fig:halo_c} (neglecting the scatter at large halo masses, which is due to the small numbers of haloes there). \tcr{Recent simulations and modelling have indicated that the mass dependence of the halo concentration can be more complicated and is not a simple powerlaw across the whole halo mass range \citep[e.g.,][]{Prada2012,Sanchez-Conde2014,Ludlow2014,Ng2014}. However, our GR simulation has too small a dynamical range to be affected by this.}

In $f(R)$ gravity, however, this is no longer true. Indeed, here we find a turning mass scale $M_{\ast}$, below which the halo $c$-$M_{200}$ relation shows a clear deviation from a single power law and becomes higher than in GR.
\tcr{We have checked this discovery by running the Amiga Halo Finder \citep[][{\sc Ahf}]{ahf}, which employes a different method to measure halo concentrations, by using the relation between the maximum circular velocity and halo mass for NFW haloes, and found good agreement.}
\tcr{We also make an additional test by fitting the halo density profiles to the Einasto formula \citep{einasto,Navarro2004}, because it is known in $\Lambda$CDM that the shape of spherically averaged halo density profiles deviates systematically (though only slightly) from the two-parameter NFW formula and can be better described by the three-parameter Einasto formula \citep{Gao2008}. The Einasto fitting is less sensitive to the radius range used in the fitting, but in the test we only use radial bins outside the \citet{Power2003} convergence radius. Again, we have found very good agreement with Fig.~\ref{fig:halo_c}. Finally, in Fig.~\ref{fig:halo_c} we have included all haloes, and in the last test we have also checked the results for relaxed haloes only, using the criteria proposed by \citet{neto}. We find that such a selection of relaxed haloes makes very little difference in the concentration-mass relation, which agrees well with the findings of \citet{Gao2008}. Since the main focus of this paper is the comparison between $f(R)$ gravity and GR, we shall not show the plots from those tests.}

\tcr{A possible reason for the difference in the concentration-mass relations of the two models studied here is the following: the turning mass scale, $M_\ast$, which itself depends time, is roughly a threshold mass for the fifth force screening in $f(R)$ gravity at each given time. Less massive haloes are unscreened and have deeper potentials than GR haloes with the same mass, which can make particles move towards the central regions and lead to  higher concentrations. The increase of $M_\ast$ with time reflects the simple fact that as time goes on more massive haloes become unscreened.}

\tcr{Similar behaviour has also been found in other modified gravity theories. As an example, 
\citet{barreira2014b, barreira2014a} find that, for models in which the strength of gravity increases rapidly in time, halos tend to be more concentrated (and vice versa). In chameleon-type theories, including $f(R)$ gravity, the screening makes the situation more complicated, but the general picture is that haloes tend to be more concentrated if the model has had an efficient screening at early times (such as F6) because, at late times when screening is `switched off', the potentials inside haloes deepen suddenly, and matter particles tend to fall towards the halo centre \citep{lz2010, zlk2011b}. 
Finally, in the phenomenological ReBEL model of \cite{rebel}, in which a scalar-mediated Yukawa-type fifth force helps in boosting the structure formation from early times, \citet{Hellwing13} notice that the halo concentration is higher for all halo masses. These authors compare the kinetic and potential energies in their virialised haloes, and find that the ratio between the two is actually smaller than in $\Lambda$CDM haloes of same masses (cf.~Fig.~11 in that paper). Even though the fifth force in the ReBEL model starts to effect from early times, the fact that it has a finite range (not longer than $1h^{-1}$Mpc in the models simulated by \citet{Hellwing13}) means that the enhanced gravity could not affect regions beyond $\sim1h^{-1}$Mpc: this is similar to the behaviour of the fifth force in F6 for our small haloes, which is possibly why the effect on the halo concentrations is also similar in the two cases.}

\tcr{Another possible reason for the different $c$-$M$ relations in F6 and GR is the different halo formation histories in the two models. As is mentioned above, haloes which form at earlier times generally have higher concentration because the mean matter density is higher when they collapse. Consider two (small) haloes of the same mass in GR and in F6: it is more likely that the latter has a larger fraction of its present-day mass assembled at later times, and thus its inner region is smaller, forms earlier and is more concentrated (in other words, a halo with mass $M_1$ in F6 is likely to have a mass $M_2<M_1$ in GR and thus have a higher concentration than a GR halo of mass $M_1$). It would be useful to disentangle the two effects affecting halo concentrations, but this is difficult because a modified gravitational force will always simultaneously affect both the halo accretion history and halo potential, except in cases where the screening is very strong inside haloes, such as in the cubic Galileon model \citep{barreira2014a}. We shall leave such a study for future work.} 

\tcr{We caution that the result for F6 may not quantitatively hold for HS $f(R)$ models with other values of $n$ or $f_{R0}$, or to other $f(R)$ or chameleon models. The complicated physics that determines the concentration implies that the $c$-$M_{200}$ relation needs to be studied on a case-by-case basis in general.}

\subsection{Halo formation histories}

\begin{figure}
\includegraphics[width=8.6cm]{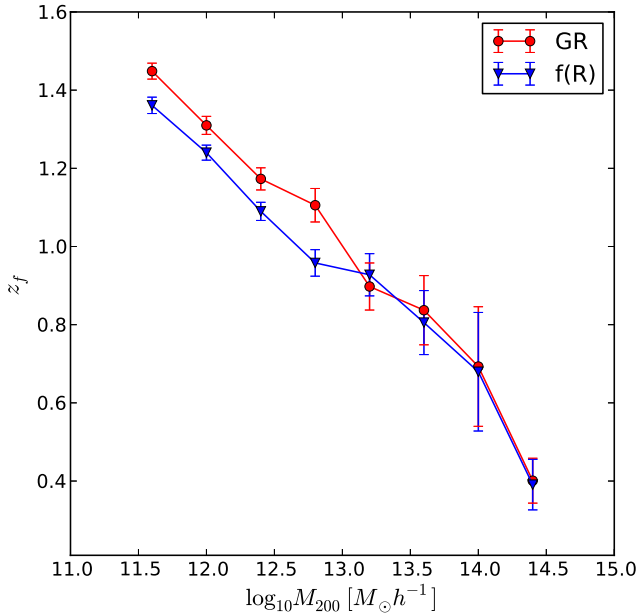}
\caption{The halo formation time $z_f$ as a function of $M_{200}$, from our GR (red circles and curve) and F6 (blue triangles and curve) simulations. Error bars are 1-$\sigma$ standard deviations.}
\label{fig:hft_M200}
\end{figure}

The formation of dark matter haloes is a complicated process, in which frequent mergers and the accretion of smaller haloes hierarchically lead to the formation of larger haloes. In this picture, large haloes form later when the environmental density is lower, and thus have lower concentrations than small haloes, as we have seen in the previous subsection.

As the gravitational force is enhanced in $f(R)$ gravity, is has been speculated that the matter clustering is stronger and as a result dark matter haloes form earlier in our $f(R)$ model than in GR. For example, a previous study by \citet{hkk2010} found that in the \tcr{ReBEL model, the Yukawa-type fifth force helps to form haloes at higher redshift than} in the standard $\Lambda$CDM model, and therefore can potentially move reionisation to earlier times as implied by CMB observations. Here, we want to study the halo formation times in our F6 simulations.

In order to follow the growth of a halo with time, we start with the halo at the present time and identify its most massive progenitor from the previous snapshot. We repeat this procedure until the halo mass is too small to be resolved anymore, and define the halo formation time as the redshift, $z_f$, at which the most massive progenitor halo has assembled half of its mass at $z=0.0$. This formation time has been widely used in the literature \citep[e.g.,][]{Lacey1993, Gao2004}, although other definitions have also been used \citep[e.g.,][]{wechsler2001}.

In Fig.~\ref{fig:hft_M200}, we plot the halo formation time $z_f$ as a function of $M_{200}$. In both models, the results agree with the above hierarchical picture that low-mass haloes form earlier. When comparing the two models, we can see that haloes more massive than $10^{13}h^{-1}M_\odot$ form at nearly the same redshift in GR and F6, showing again that the chameleon mechanism works efficiently for these haloes to suppress the effects of modified gravity. Less massive haloes, on the other hand, form slightly {\it earlier} in GR than in F6. 
This result seems to disagree with the general pattern found in \citet{hkk2010} in the ReBEL model, and seems in contrast to the naive expectation that the enhanced gravitational force in F6 boosts the hierarchical structure formation.

\begin{figure*}
\includegraphics[width=17.8cm]{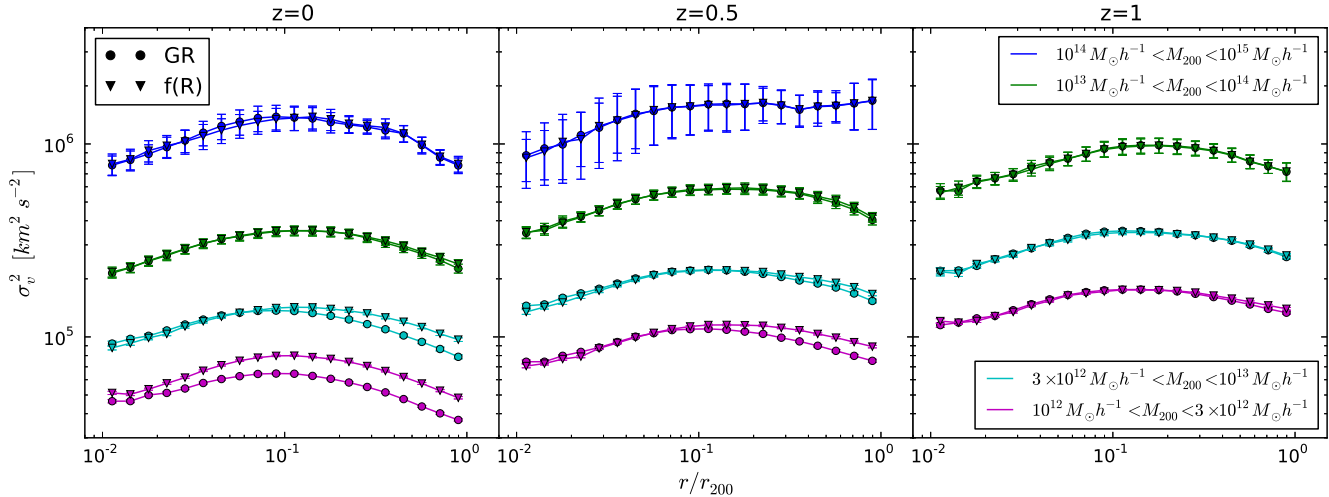}
\caption{The stacked velocity dispersion profiles from our $f(R)$ (filled triangles) and GR (filled circles) simulations, at three redshifts: $z=0$ (left panel), $z=0.5$ (middle panel) and $z=1$ (right panel). We show the results for \tcr{four} bins of the host halo mass increasing from the bottom curves to the top curves, and indicated in the legend \tcr{(we have divided the mass bin $10^{12}\sim10^{13}h^{-1}M_\odot$ into two sub-bins because this is the only bin which shows difference between F6 and GR)}. The solid curves simply connect the symbols, and the error bars show the 1-$\sigma$ scatter around the mean.}
\label{fig:sigma_v}
\end{figure*}

To explain this behaviour, we need to again carefully examine the subtle differences between different models. In ReBEL, there is a Yukawa-type fifth force between particles, whose strength decays with distance but does not change in time. This implies that the fifth force starts to boost structure formation from early times, resulting in haloes forming earlier. In the case of F6, gravity is suppressed at redshift $z\gtrsim1$, and even at $z\lesssim1$ it is only enhanced for smaller haloes. This means that:
\newline\indent(i) the formation history of very massive haloes $({\rm e.g.,~}M\gtrsim10^{13}h^{-1}M_\odot)$ does not see the effect of an enhanced gravity, as we have seen above;
\newline\indent(ii) less massive haloes evolve in a similar manner as in GR at $z\gtrsim1$, but grow more rapidly at $z\lesssim1$, and as such they are more massive than their GR counterparts at present. As $z_f$ is defined to be the time when a halo has gained half of its current mass (denoted by $M_{1/2}$), the halo would have a larger $M_{1/2}$ in F6 than in GR. But small haloes typically grow to $M_{1/2}$ at $z\gtrsim1$, before when there is little difference between GR and F6, and so it takes the halo longer to acquire a mass of $M_{1/2}$ in F6 than in GR, which means that the halo forms {\it later} in $f(R)$ gravity. This, of course, is purely a result of the definition of $z_f$, and does {\it not} imply that matter clusters more slowly in F6.

Therefore, like the concentration, the halo formation time also depends sensitively on the nature of gravity. Even for two models in both of which gravity is enhanced, the behaviour of $c\left(M_{200}\right)$ or $z_f\left(M_{200}\right)$ can be qualitatively different. For this same reason, the results for $z_f$ for F6 can not be generalised to other variants of the HS $f(R)$ model or other $f(R)$ models without careful tests.

\subsection{Halo velocity dispersion profiles}

Before leaving this section, we study the velocity dispersion profile in our simulations, which is defined as
\begin{eqnarray}\label{eq:sigam_v}
\sigma_v(r)^2\equiv\frac{1}{\Delta N_p}\sum_{i\in\Delta r} (\vec{v}_i-\vec{v}_h)^2,
\end{eqnarray}
in which $i\in\Delta r$ means that particle $i$ sits in a spherical shell from radius $r-\Delta r/2$ to $r+\Delta r/2$, and $N_p$ is the number of particles within this shell. $\vec{v}_i$ and $\vec{v}_h$ are the particle and host halo velocities respectively, and the latter is calculated as the average of the velocities of the 25\% most bound particles in the host halo. The halo velocity dispersion is a more direct characterisation of the potential inside a halo; it is determined by the dynamical \citep{fabian,zlk2011} or effective \citep{hhlg2015} mass of a halo, and is enhanced by the modified force for unscreened haloes \citep{lzlk2013,hhlg2015}.

In Fig.~\ref{fig:sigma_v} we show the velocity dispersion profiles measured from our simulations at $z=0.0$ (left), $0.5$ (middle) and $1.0$ (right). Thanks to the chameleon screening, the difference between the two models for haloes more massive than $\sim10^{13}h^{-1}M_{\odot}$ is almost undetectable. Haloes in the mass range $10^{12}-10^{13}h^{-1}M_\odot$ can have significantly higher velocity dispersion in $f(R)$ gravity than in GR, and the deviation increases with the distance from the halo centre, since the screening in $f(R)$ gravity is relatively weak inside small halos, particularly in their outer regions in which matter density is low. We also notice that the enhancement of velocity dispersion is weaker at earlier times, due to stronger chameleon screening and less time for the fifth force to take effect.

The result confirms that particles bound in unscreened haloes have higher kinetic energy to balance the extra potential produced by the fifth force. This implies that measurements of galaxy velocity dispersions in galaxy groups, such as the Local Group, may not be able to give reliable estimates of the true masses of the systems. For example, to use such measurements to find the underlying mass requires a good understanding of the screening, which in turn requires an accurate knowledge of the true mass (as well as the environmental effects). Therefore, a trial-and-error procedure would be needed to improve the mass estimation iteratively from some initial guess, and each iteration needs to be calibrated by high-resolution simulations which take into account the full environmental effects and other complexities such as irregular shapes of haloes and distributions of their massive satellites.

On the other hand, if we indeed live in an unscreened region in $f(R)$ gravity, but choose to interpret our measurements of galaxy velocity dispersions in the incorrect framework of GR, then the estimated mass will be biased high compared with its true value. We will briefly mention one of its implications below. In any case, it is clear that $f(R)$ gravity would make the already uncertain estimates of the Milky Way mass even more complicated.

\section{Properties of substructures}
\label{sect:subhalos}

In the previous section we analysed the simulation results of various halo properties in our F6 and GR simulations. In this section, we turn our attention to the properties of subhaloes in these models.

In hierarchical structure formation, halo merger events leave plenty of remnant structures that survive as subhaloes in the descendent haloes. As galaxies form inside haloes and migrate with them, subhaloes then exist as the host sites of satellite galaxies in galaxy groups and clusters. The properties of subhaloes and their evolution history (i.e., the subhalo merger tree) provide the backbone for models of galaxy formation~\citep[see, e.g.,][for a review]{Baugh06}. The abundance and distribution of subhaloes also has important implications for the indirect detection of dark matter, for example by boosting the dark matter annihilation signal~\citep[e.g.,][]{Fermi-Gao,Fermi-Han}.

The fact that subhaloes form through hierachical mergers can also be utilised to identify them. Here we will use the tracking subhalo finder Hierarchical Bound-Tracing \citep[][HBT]{HBT} to identify subhaloes. Starting from isolated haloes at an earlier snapshot, HBT identifies their descendents at subsequent snapshots and keeps track of their growth. As soon as two haloes merge, HBT starts to track the self-bound part of the smaller progenitor as a subhalo in each subsequent snapshot. With a single walk through all the snapshots, all the subhaloes formed from halo mergers can be identified in this way. Such a unique tracking algorithm enables HBT to largely avoid the resolution problem suffered by configuration space subhalo finders~\citep{Muldrew, HBT, Onions}. By construction, HBT also produces clean and self-consistent merger trees that naturally avoid subtle defects such as missing links and central-satellite swaps common to many other tree builders~\citep{Suss,Suss2}.

\subsection{Subhalo mass functions}

\begin{figure*}
\includegraphics[width=18cm]{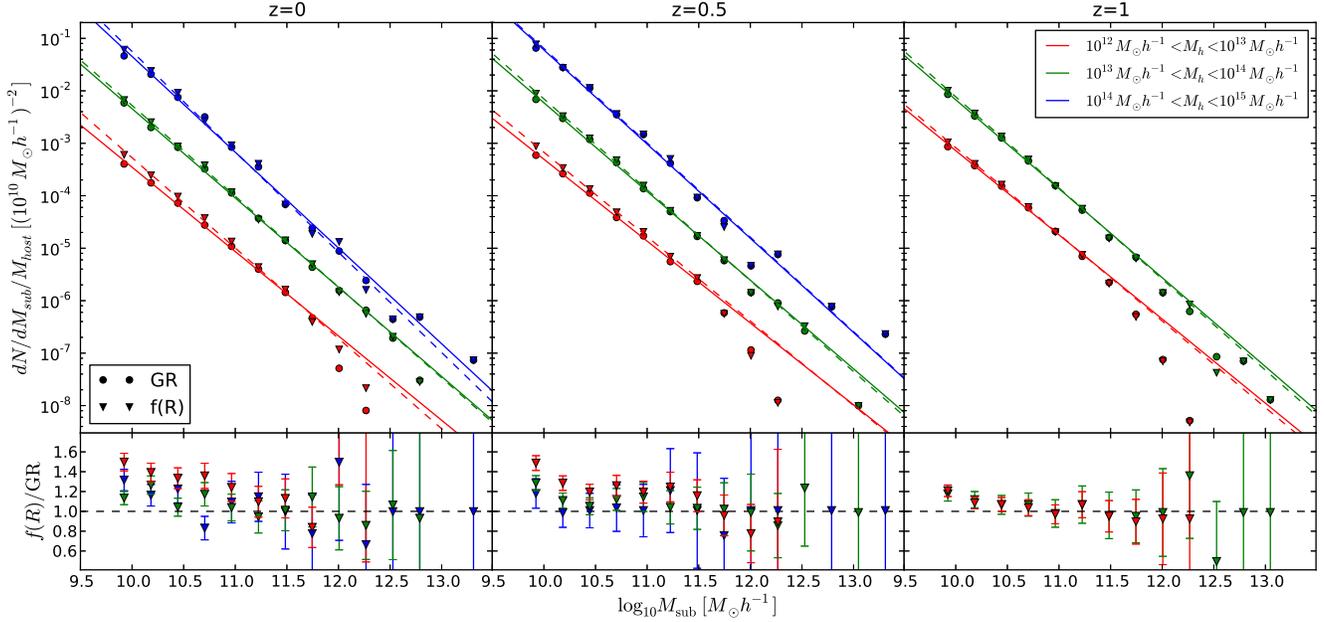}
\caption{The differential subhalo mass function ${\rm d}N/{\rm d}M_{\rm sub}$ as a function of the subhalo mass $M_{\rm sub}$, at three redshifts -- $z=0.0$ (left panel), $z=0.5$ (middle panel) and $z=1.0$ (right panel). The lower subpanels show the ratio between the F6 and GR results, and the binning scheme of host halo masses is indicated by different colours as shown in the legends (note that the highest mass bin does not exist in the right panels because at $z=1$ haloes more massive than $10^{14}h^{-1}M_{\odot}$ does not exist in great numbers). The error bars (only shown in the lower subpanels for clarity) are $1$-$\sigma$ standard deviations in each subhalo mass bin similarly as in the halo mass function plots. For guide eyes, we also plot the power-law fitting results of the subhalo mass functions using lines with the same colours (solid for GR and dashed for F6).
}
\label{fig:shmf}
\end{figure*}

Similar to haloes, the abundance of subhaloes can be described by a subhalo mass function (SHMF). The SHMF is known to depend on the size of their host haloes \citep{Gao2004, vandenBosch} in $\Lambda$CDM simulations, but is close to a universal power law function of the subhalo mass, $M_{\rm sub}$, when normalised by the host halo mass, $M_{\rm host}$.

In Fig.~\ref{fig:shmf} we plot the SHMF for three bins of host halo mass, $10^{12}\sim10^{13}h^{-1}M_\odot$, $10^{13}\sim10^{14}h^{-1}M_\odot$, $10^{13}\sim10^{14}h^{-1}M_\odot$ (see the legends), at three redshifts, $z=0$ (left), $0.5$ (middle) and $1$ (right).  For clarity, the results for the highest (lowest) mass bin are shifted upwards (downwards) by a decade. A quick visual inspection of Fig.~\ref{fig:shmf} indicates that the power-law relation holds true for the $\Lambda$CDM (circles and solid lines) and F6 (triangles and dashed lines) simulations as well, though the slope has a weak dependence on the host halo mass (lower for low-mass host haloes). To check this result, we tested HBT on a simulation using a different N-body code \citep[described in][]{js2002} and found the same tendency. We also tested our simulations using the {\sc Rockstar} code \citep{rockstar} to identify subhaloes, but did not notice any dependence of this slope on the host halo mass. Therefore, we conclude that this is likely due to  the subhalo finding algorithm we use, which finds more massive and extended subhaloes than some other algorithms \citep{HBT}. 
\tcred{We note that, even though the SHMF from HBT has a lower slope than the result from {\sc Rockstar}, it is consistently higher for the range of subhalo mass shown in Fig.~\ref{fig:shmf}.}
Because we are mainly interested in the relative differences between models in this paper, we will leave a more detailed comparison of different algorithms to a future separate work \tcr{and not show a plot for the comparison}.

From Fig.~\ref{fig:shmf} we find that the difference between F6 and GR is smaller for more massive host haloes and at earlier times, because in both cases the chameleon screening is more efficient and effects of modified gravity more strongly suppressed. Differences between the two models also tend to be larger for small subhaloes, with F6 predicting $20\sim50\%$ more subhaloes with $M_{\rm sub}$ between $10^{11}$ and $10^{10}h^{-1}M_\odot$ than GR in host haloes of mass $10^{12}\sim10^{13}h^{-1}M_\odot$. This implies that the enhanced gravity in the $f(R)$ model studied here can help produce a substantially higher abundance of substructures in Milky-Way-sized dark matter haloes. We will discuss the implication of this in the context of Milky Way satellite abundances below when discussing the subhalo velocity function.

Note that an enhanced gravity will not only boost the clustering of matter and formation of subhaloes, but can also increase the stripping of matter from subhaloes inside haloes (and thus decrease subhalo masses). Our results above suggest that the latter effect is subdominant.

\subsection{Subhalo spatial distributions}

\begin{figure}
\includegraphics[width=9.2cm]{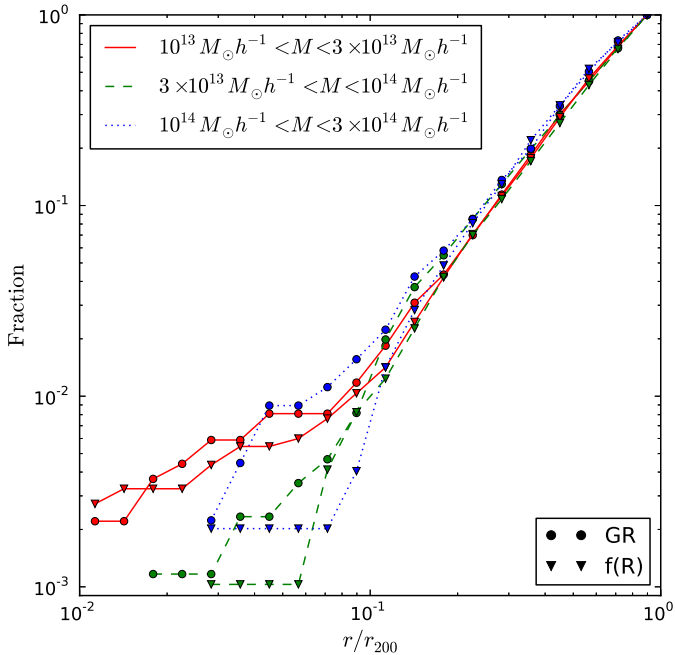}
\caption{The cumulative radial number distributions for subhalos, for three mass bins of their host haloes as indicated by the legends. The vertical axis is the fraction of subhaloes within $r/r_{200}$, and we show the results for both F6 (triangles) and GR (circles) at $z=0$.}
\label{fig:sh_dist}
\end{figure}

Next we focus on the spatial distribution of subhaloes inside their host haloes. Naturally, one expects this distribution to depend on the nature of gravity,
though this dependence can be weakened by the chameleon screening by the host haloes in $f(R)$ gravity.

\begin{figure*}
\includegraphics[width=17.9cm]{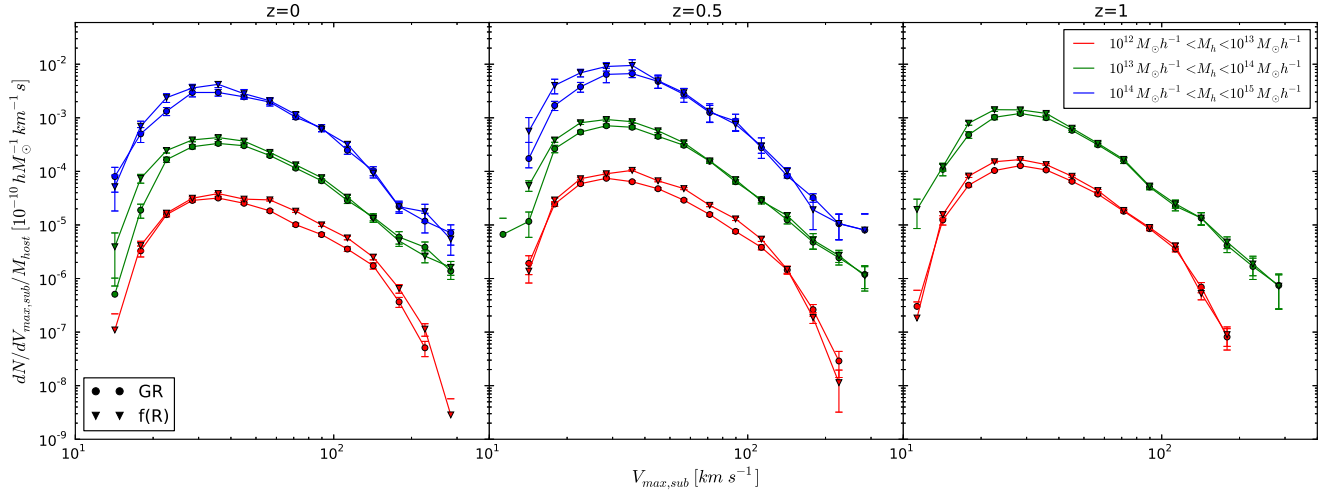}
\caption{The differential subhalo velocity functions from our $f(R)$ (filled triangles) and GR (filled circles) simulations, at three redshifts: $z=0$ (left panel), $z=0.5$ (middle panel) and $z=1$ (right panel). We show the results for three bins of the host halo mass increasing form the bottom curves to the top curves, and indicated in the legend; the solid curves simply connect the symbols. Note that the number of subhaloes in each bin is normalised by the total mass of the host halo. The results for the highest (lowest) host halo mass bin are artificially shifted upwards (downwards) by a decade to make the plot clearer. Error bars show the 1-$\sigma$ scatter around the mean.}
\label{fig:shvf}
\end{figure*}

\citet{Gao2004} showed that the spatial distribution of subhaloes does not have a significant dependence on their host halo masses. In our $\Lambda$CDM simulations we have found the same result, as shown in Fig.~\ref{fig:sh_dist}, in which we plot the cumulative radial number distributions of subhaloes as circles for $\Lambda$CDM. We show in different colours the results for three mass bins of host haloes, all at $z=0$, which agree well with each other.

To see the effect of $f(R)$ gravity, we also plot the corresponding results from the F6 simulation in Fig.~\ref{fig:sh_dist} using triangles. There is very little difference from the GR results, possibly because of the efficient screening. Notice that here we have only shown results for host haloes more massive than $10^{13}h^{-1}M_\odot$, in which the modified gravity effects are strongly suppressed as we have seen above. The results for smaller host haloes are not shown since they are noisier due to resolution limitations.

\subsection{Subhalo velocity function}

Subhaloes reside in the high-density environments within their host haloes, and experience constant tidal stripping, which strips mass from their outer parts. Their mass could change significantly during their evolution. Therefore, in the literature people often use the maximum circular velocity $V_{\rm max}$ instead, because it depends primarily on the inner part of a (sub)halo.

Following \citet{Gao2004}, in Fig.~\ref{fig:shvf} we plot the differential abundance of subhaloes as a function of $V_{\rm max}$, also known as the subhalo velocity function (SHVF). The $\Lambda$CDM results in the range of $30~{\rm km/s}\lesssim V_{\rm max}\lesssim200~{\rm km/s}$ are well described by a universal power-law function, in agreement with findings in the literature \citep[e.g.,][]{Gao2004,dooley} (note that we have shifted the curves for different host halo mass bins for clarity), and drop off at small (large) $V_{\rm max}$ is due to the resolution limit (finite box size).

In $f(R)$ gravity, the qualitative behaviour of the SHVF is similar, but the enhanced gravity leads to quantitative differences. For more massive host haloes, the difference is most significant at small $V_{\rm max}$, which correspond to smaller subhaloes that are less screened and therefore have formed in higher abundances; in contrast, larger subhaloes, with larger $V_{\rm max}$, are better screened and so their abundances do not change significantly from the GR predictions.  For less massive host haloes, there is a noticeable boost in the subhalo abundance even for large subhalo $V_{\rm max}$, since the host haloes have become less screened since earlier times and substructures have more time to grow there. This dependence on host halo mass in principle implies a deviation from a universal SHVF, although the effect we see in Fig.~\ref{fig:shvf} is fairly weak.

The enhanced SHVF at $V_{\rm max}\gtrsim30~{\rm km/s}$ for host haloes with mass of $\sim10^{12}h^{-1}M_\odot$ seems to suggest that the missing massive satellite problem of the Milky Way galaxy is worse in $f(R)$ gravity, since in the latter the {observed} number of dwarf galaxies remains the same while the {theoretically predicted} number of massive subhaloes is larger. \citet{Wang2012} argue that the missing satellite problem is not serious enough to motivate a revision to the $\Lambda$CDM paradigm, but what we saw above in Fig.~\ref{fig:shvf} certainly seems to make $f(R)$ gravity disfavoured. However, there are complicated issues which preclude a definite conclusion. For example, most measurements of the Milky Way halo actually predict its dynamical mass, which can be 1/3 heavier than the true mass in $f(R)$ gravity -- hence, with a given rotation speed of the Milky Way disk, the actual mass of the halo could be smaller than what we currently think. Also, galaxy formation can also be different in $f(R)$ gravity, so that the way in which galaxies populate massive subhaloes might be different, making a direct comparison with $\Lambda$CDM even harder. We will leave detailed studies of these issues to future works.

\section{Discussions and conclusions}
\label{sect:summary}

To briefly summarise, in this paper we have employed a very high-resolution simulation to study the properties of dark matter haloes and subhaloes of a $f(R)$ gravity model. This model is a variant of that proposed by \citet{hs2007}, with parameters $n=1$ and $|f_{R0}|=10^{-6}$. We argue that this is a borderline model that should be studied as a first step towards a more rigorous constraint on $f_{R0}$ combining cosmological and astrophysical observations. We regard this as a realistic model which is not yet apparently ruled out by current data.

The simulations we use in our analyses have $512^3$ particles in a cubic box of size $L_{\rm box}=64h^{-1}$Mpc, with a background cosmology chosen to be that of the best-fit WMAP9. This cosmology is more updated and more realistic than those of the previous $f(R)$ simulations conducted by us \citep[e.g.,][]{LiPS2013}, and the resolution here is also significantly higher, making it possible to study subhaloes in detail. Our halo catalogue is constructed using a standard friend-of-friend linking method, and the subhaloes were found using the HBT algorithm of \citet{HBT}.

Due to the efficient chameleon screening, this $f(R)$ model shows small deviations from $\Lambda$CDM in general. For example, the halo mass function shows at most $\sim20\%$ enhancement compared with the $\Lambda$CDM result between $z=0$ and $z=1$, with the deviation propagating to more massive haloes as time passes, in agreement with the semi-analytical predictions of \citet{le2012}. The dark matter distribution inside halos is almost identical in this $f(R)$ model as in $\Lambda$CDM for haloes more massive than $\sim10^{13}h^{-1}M_\odot$, again due to the chameleon screening; however, for smaller haloes, the screening is less efficient, which results in a deepening of the total potential and subsequently a steepening of the density profile. As a result, the halo concentration-mass relation is enhanced for such low-mass haloes and can no longer be described by a simple power law (as for $\Lambda$CDM). The stronger gravitational force in this $f(R)$ model also enhances the growth of small haloes, but mainly at late times and as a result the halo formation time (i.e., the time by which a halo has gained half of its present-day mass) is actually later than in $\Lambda$CDM. We stress that these conclusions hold only for this specific $f(R)$ model, and there is evidence suggesting that other models could behave qualitatively differently because of the complicated behaviour of gravity. We also notice enhanced halo velocity profiles in this $f(R)$ model, confirming various previous work \citep[e.g.,][]{ctl2015,glmw2014,hhlg2015}.

The stronger gravity also helps to produce more substructures, mainly in host haloes less massive than $\sim10^{13}h^{-1}M_\odot$ because of the weaker screening therein, and for subhaloes less massive than $10^{12}h^{-1}M_\odot$. We find that Milky Way-sized haloes could host up to $20\sim50\%$ more subhaloes in the mass range $10^{10}\sim10^{11}$ $h^{-1}M_\odot$ in the studied $f(R)$ model than in $\Lambda$CDM. The subhalo mass function can be fitted using a simple power law, as in $\Lambda$CDM, but with different parameters. We do not find a noticeable difference in the radial distribution of subhaloes inside their host haloes between the two models, though. The higher abundance of substructures is confirmed in the subhalo velocity functions, which seems to make the missing satellite problem of the Milky Way worse. However, we stress that there are caveats in interpreting the result at its face, due to the further complexities in observationally determining halo mass in the context of modified gravity.

Overall, we find that halo and subhalo properties of this borderline $f(R)$ model are close to the $\Lambda$CDM predictions for massive haloes, confirming previous results that this model is difficult to distinguish from $\Lambda$CDM using cosmological observations. However, a substantial deviation might be found in less massive haloes such as that of our Milky Way, which is in agreement with the findings of previous low resolution simulations. This indicates that the dynamics of systems such as the Local Group can be sensitive to modifications of gravity of this kind and strength. This should be a focus of further studies in the future, following the recent progress in zoom simulations made by \citet{ctl2015}.

As mentioned above, this is a first step of a more detailed study of this borderline model, and here we have not touched the topic of astrophysical constraints, which is much more complicated. Studies of \citet[][]{bvj2013} and \citet{vsbn2014} have demonstrated the potential of using astrophysical systems to improve the constraints on $f_{R0}$. It would be useful to have a better understanding of the impact that environmental screening could have on those constraints. As in $f(R)$ models the local behaviour of gravity usually depends on its environment at much larger scales, high-resolution or zoom simulations are important for calibrating the interpretation of astrophysical observations. They are also important because they can provide more realistic quantifications of the environments for stellar evolution, which depends on the nature of gravitation sensitively \citep{dlss2012}.

Obviously, improved constraints may or may not rule out this $f(R)$ gravity model. However, with the progress in both numerical simulations and theoretical modelling, we are on a path towards better understanding. In such a sense, we are currently in the state of {\it liminality}\footnote{In the literature of Education, {\it liminality} refers to the suspended state of partial understanding, which is unsettling and lacks of authenticity. Such a state can be crossed and re-crossed many times, and its final crossing marks the opening of a world of new knowledge.}, and much effort is still in need to pass it.

\section*{Acknowledgments}

{We thank Alexandre Barreira, Carlton Baugh, Sownak Bose, Jianhua He and Matthieu Schaller for helpful discussions and/or comments on the draft}. DS acknowledges support by the European Research Council Starting Grant (DEGAS-259586) of Peder Norberg and Euclid implementation phase (ST/KP3305/1). BL is supported by the Royal Astronomical Society and Durham University.
This work was supported by the UK Science and Technology Facilities Council (STFC) Grant no.~ST/L00075X/1. This work used the DiRAC Data Centric System at Durham University, operated by the Institute for Computational Cosmology on behalf of the STFC DiRAC HPC Facility (http://www.dirac.ac.uk). This equipment was funded by BIS National E-infrastructure capital grant ST/K00042X/1, STFC capital grant ST/H008519/1, STFC DiRAC Operations grant ST/K003267/1 and Durham University. DiRAC is part of the National E-infrastructure.

\bibliographystyle{mn2e}
\bibliography{fr_subhalos}

\label{lastpage}

\end{document}